%% file: main.tex
\documentclass[twocolumn]{aastex63}

\input{def}

\received{June 1, 2019}
\revised{January 10, 2019}
\accepted{\today}
\submitjournal{ApJL}

\shorttitle{SN~2011fe}
\shortauthors{M. A. Tucker et al.}
\graphicspath{{./}{figures/}}

\begin{document}

\title{A Rapid Ionization Change in the Nebular-Phase Spectra of the Type Ia SN 2011fe}

\correspondingauthor{Michael Tucker}
\email{tuckerma95@gmail.com}

\author[0000-0002-2471-8442]{M. A. Tucker}
\altaffiliation{DOE CSGF Fellow}
\affiliation{Institute for Astronomy,
University of Hawai`i at Manoa,
2680 Woodlawn Dr., Honolulu, HI, USA}

\author{C. Ashall}
\affiliation{Institute for Astronomy,
University of Hawai`i at Manoa,
2680 Woodlawn Dr., Honolulu, HI, USA}

\author{B. J. Shappee}
\affiliation{Institute for Astronomy,
University of Hawai`i at Manoa,
2680 Woodlawn Dr., Honolulu, HI, USA}

\author{C. S. Kochanek}
\affiliation{Department of Astronomy, The Ohio State University, 140 West 18th Avenue, Columbus, OH 43210, USA}
\affiliation{Center for Cosmology and AstroParticle Physics, The Ohio State University, 191 W. Woodruff Ave., Columbus, OH 43210, USA}

\author{K. Z. Stanek}
\affiliation{Department of Astronomy, The Ohio State University, 140 West 18th Avenue, Columbus, OH 43210, USA}
\affiliation{Center for Cosmology and AstroParticle Physics, The Ohio State University, 191 W. Woodruff Ave., Columbus, OH 43210, USA}

\author{P. Garnavich}
\affiliation{Physics Department, University of Notre Dame, Notre Dame, IN 46556, USA}



\begin{abstract}

We present three new spectra of the nearby Type Ia supernova (SN Ia) 2011fe covering $\approx 480-850$~days after maximum light and show that the ejecta undergoes a rapid ionization shift at $\sim 500$~days after explosion. The prominent \fFe{3} emission lines at $\approx 4600$~\AAA are replaced with \ion{Fe}{1}+\ion{Fe}{2} blends at $\sim 4400$~\AAA and $\sim 5400$~\AAA. The $\approx 7300$~\AAA feature, which is produced by \fFe{2}+\fNi{2} at $\lesssim 400$~days after explosion, is replaced by broad ($\approx \pm 15\,000$~\kms) symmetric \CaII emission. Models predict this ionization transition occurring $\sim 100$~days later than what is observed, which we attribute to clumping in the ejecta. Finally, we use the nebular-phase spectra to test several proposed progenitor scenarios for SN~2011fe. Non-detections of H and He exclude nearby non-degenerate companions, [\ion{O}{1}] non-detections disfavor the violent merger of two white dwarfs, and the symmetric emission-line profiles favor a symmetric explosion.

\end{abstract}



\section{Introduction}\label{sec:intro}

Type Ia supernovae (SNe Ia) are the thermonuclear explosions of carbon/oxygen (C/O) white dwarfs (WDs; \citealp{hoyle1960}) and produce the majority of iron-group elements in the Universe \citep[e.g., ][]{iwamoto1999}. These stellar explosions attracted much attention due their use as standardizable candles \citep{phillips1993} but we still lack a genuine understanding of how and why some WDs explode as \sneia. 

\name was discovered $\approx 11$~hours after explosion by the Palomar Transient Facility \citep[PTF; ][]{law2009} in M101 at a mere $6.4~\rm{Mpc}$ \citep[e.g., ][]{shappee2011} and became the brightest \snia in several decades. Additionally, \name is the quintessential \snia \citep{pereira2013}, making it an ideal object for testing \snia progenitor and explosion models \citep[e.g., ][]{ropke2012}. This is especially important for nebular-phase observations when the ejecta becomes optically-thin to optical and NIR photons to provide a direct view to the inner ejecta. 

Several studies have already analyzed spectra of \name as it transitioned into the nebular phase. \citet{shappee2013} and \citet{lundqvist2015} searched for \Ha emission indicative of a nearby donor star at the time of explosion and found none. \citet{mcclelland2013} showed that the nebular-phase mid-infrared (MIR) decay is correlated with the nucleosynthetic yield. \citet{mazzali2015} modeled a suite of photospheric- and nebular-phase spectra and claimed the existence of a stable iron core, although this remains debated \citep[e.g., ][]{botyanszki2017}. Finally, \citet{taubenberger2015} and \citet{graham2015b} analyzed spectra acquired $\approx 1000$~days after maximum light, the latest spectra of a \snia to-date. They found that the emission characteristics were drastically different from the previous spectra at $\approx 300$~days. Models by \citet{fransson2015} show that the 1000-day spectrum is dominated by \pFe{1} emission lines with minor contributions from \pFe{2} and Ca. However, the spectroscopic evolution between $\sim 400-1000$~days after explosion remains poorly understood. \citet{friesen2017} briefly attempted to model the +576-day spectrum from \citet{graham2015b} obtained by the Berkeley SuperNova Identification Program \citep[BSNIP; ][]{silverman2012} but the results were unsatisfactory and they focused on modeling earlier-phase spectra. 

In this Letter we analyze new nebular-phase spectra of \name acquired $\approx 480-850$~days after peak-light. Details about the data reduction and calibration are provided in \S\ref{sec:data}. \S\ref{sec:results} presents the temporal evolution including a distinct change in the iron blends at $4000-6000$~\AAA and the appearance of \CaII. Constraints on the progenitor system and explosion mechanism are described in \S\ref{sec:progenitor}. Finally, we discuss the rapid ionization change in the context of \snia models and the use of \CaII as an ejecta diagnostic in \S\ref{sec:discuss}. For our analysis we adopt the time of maximum-light in the $B$-band of $t_{\rm{max}} = \rm{MJD}~55813.98\pm0.03$ from \citet{zhang2016}. This is interchangeable with the time of explosion using the $\approx 18$-day rise-time of \name from \citet{pereira2013}. We use the distance to M101 of $6.4~\rm{Mpc}$ \citep{shappee2011} as in previous studies and all phases are given relative to \tmax in the SN rest-frame.

\section{New and Archival Spectra}\label{sec:data}

\input{lbtinfo}

We obtained new spectroscopic observations of \name $\approx 480-850$~days after maximum light with the Multi-Object Double Spectrograph (MODS; \citealp{pogge2010}) on the Large Binocular Telescope (LBT). Details about the spectroscopic data reduction are described in \citet{shappee2013}. In brief, each frame is bias-subtracted and flat-fielded before detecting and removing cosmic rays with \textsc{lacosmic} \citep{vandokkum2001}. Then, we extract the 1D spectrum with \textsc{iraf} and derive the wavelength calibration with arc-lamp frames. Finally, each spectrum is corrected for instrumental response and placed on a \textit{relative} flux scale with standard star observations. Table \ref{tab:lbtinfo} summarizes the observations. 

Archival spectra from \citet{shappee2013, graham2015a, graham2015b, taubenberger2015, mazzali2015} and \citet{zhang2016} are included in our analysis to increase the temporal coverage. We also use photometric observations from \citet{munari2013}, \citet{tsvetkov2013}, \citet{zhang2016}, and \citet{shappee2017} to place the spectra on an absolute flux scale. We restrict the photometry to the $BVR$ filters because 1) our spectral analysis is focused on the wavelength range covered by these filters, and 2) data from these filters are available for the spectral epochs used in our analysis (e.g., Fig. 1 from \citealp{tucker2021c}). We fit the light curves with low-order splines over the range of $200-1000$~days. We added Gaussian deviates of the estimated noise and refit the data multiple times to estimate the uncertainties in the spline model. To account for minor differences in the filter throughputs between the photometric setups we include a conservative $0.05$~mag ($\sim 5\%$) systematic uncertainty in the final spline fit. 

To flux-calibrate each spectrum, we interpolate the $BVR$ magnitudes and uncertainties using the spline fits. The synthetic $BVR$ filter magnitude is calculated from the spectrum using Eq. 7 from \citet{fukugita1996} and the spectrum is linearly scaled to match the observed $BVR$ magnitudes. Nebular-phase photometry in other optical filters ($gri$, \citealp{firth2015, kerzendorf2014, kerzendorf2017}) is used to check that the spectroscopic flux calibration is correct given the measurement measurement uncertainties. 

\section{Temporal Evolution}\label{sec:results}

\begin{figure*}
    \centering
    \includegraphics[width=\linewidth]{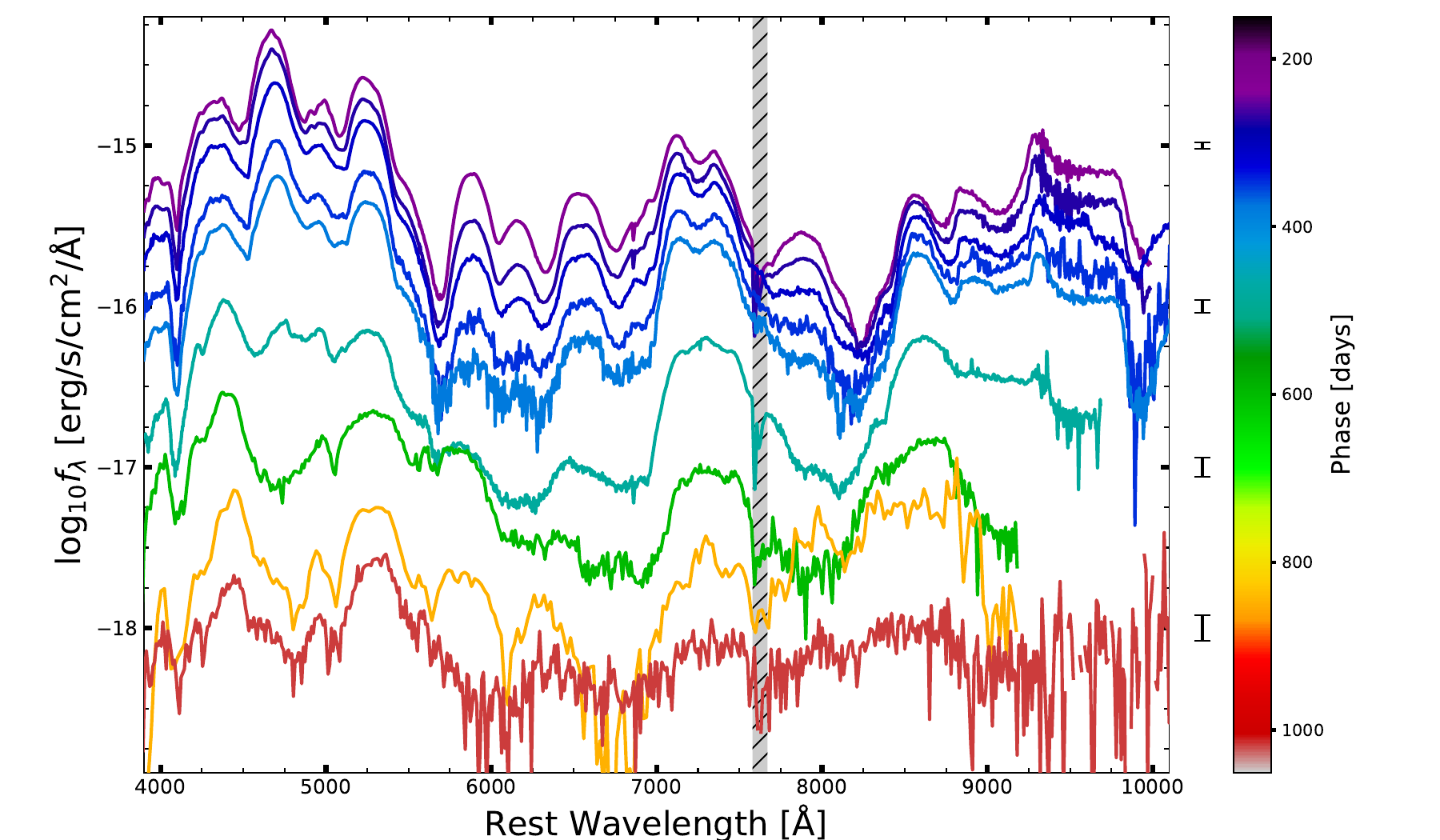}
    \caption{Flux-calibrated spectroscopic evolution of \name at $>200$ days after maximum light. Error bars along the right axis represent typical flux-calibration uncertainties after the $BVR$ spline fits (see \S\ref{sec:data}). The gray hatched region marks the O$_2$ A-band telluric feature.}
    \label{fig:fluxcal_spectra}
\end{figure*}

Line identification at $\lesssim 300~$days after maximum light is well-established due to the increasing number of nebular spectra obtained in recent years \citep[e.g., ][]{graham2017, maguire2018, sand2018, tucker2020} and improvements in modeling these phases \citep[e.g., ][]{mazzali2015, flors2018, wilk2018, flors2020, wilk2020, shingles2020, polin2021}. Optical spectra $\sim 200-300$~days after \tmax are dominated by \fFe{3} with contributions from \fFe{2}, \fNi{2}, and \fCo{3}. Fig. \ref{fig:fluxcal_spectra} shows the flux-calibrated spectral evolution at $>200$~days after \tmax and Fig. \ref{fig:normspex} shows the normalized spectra at various epochs. We discuss the evolution of the $4000-6000$~\AAA and $\approx 7300$~\AAA regions in \S\ref{subsec:results.feblend} and \S\ref{subsec:results.caii}, respectively.

\subsection{The Fe-dominated $4000-6000$~\AAA Region}\label{subsec:results.feblend}

\begin{figure*}
    \centering
    \includegraphics[width=\linewidth]{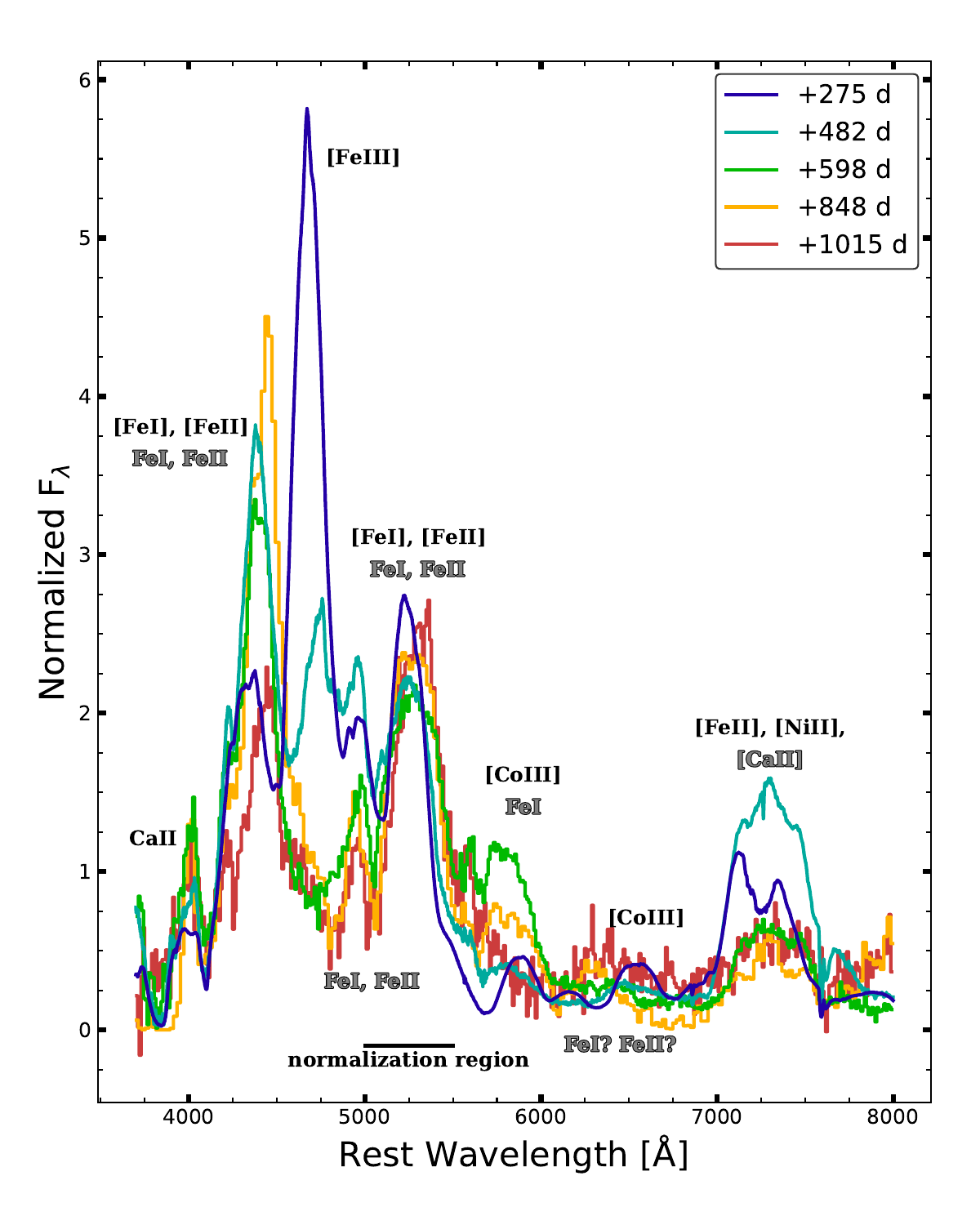}
    \caption{Spectroscopic comparison at various epochs using the same color scheme as Fig. \ref{fig:fluxcal_spectra} but normalized by the mean flux near $5300$~\AAA. Dominant line identifications before and after the distinct ionization transition at $\sim 500$~days are given in black and gray, respectively. Lines with a `?' indicate that the line's contribution to the observed spectral feature is unclear. Fe emission lines after the ionization transition are likely blends of forbidden and permitted transitions \citep{fransson2015} but we omit the bracket notation for visual clarity.}
    \label{fig:normspex}
\end{figure*}

The $4000-6000~\rm\AA$ wavelength range is mainly comprised of forbidden iron emission lines \citep[e.g., ][]{mazzali2015, wilk2020}. The strongest feature in nebular spectra is typically the peak at $\sim 4600$~\AAA observed in almost all nebular \sneia spectra and attributed to a blend of \fFe{3} transitions \citep[e.g., ][]{graham2017, maguire2018, tucker2020}. Figs. \ref{fig:fluxcal_spectra} and \ref{fig:normspex} show that this feature dominates the spectrum until $\sim400$~days after \tmax. It then disappears over the subsequent 200~days leaving two adjacent peaks at $\sim 4400~\rm\AA$ and $\sim 5300$~\AA. \citet{fransson2015} show that these features are dominated \pFe{1} with minor contributions from \pFe{2}.

Our new spectra provide the first evidence for a sharp transition in the ejecta of \sneia at nebular phases. The +480-day spectrum shows a steep drop in the \fFe{3} blend at $\approx 4600$~\AAA and the feature has disappeared by the +600-day spectrum. The adjacent spectral features at $\sim 4400$~\AAA and $\sim 5400$~\AAA remain strong in all of the spectra. However, these features shift and change compared to the $<400$-day spectra. This is likely due to the transition from \pFe{2} to \pFe{1} emission.

\subsection{The $7300$\AAA Feature}\label{subsec:results.caii}

\begin{figure}
    \centering
    \includegraphics[width=\linewidth]{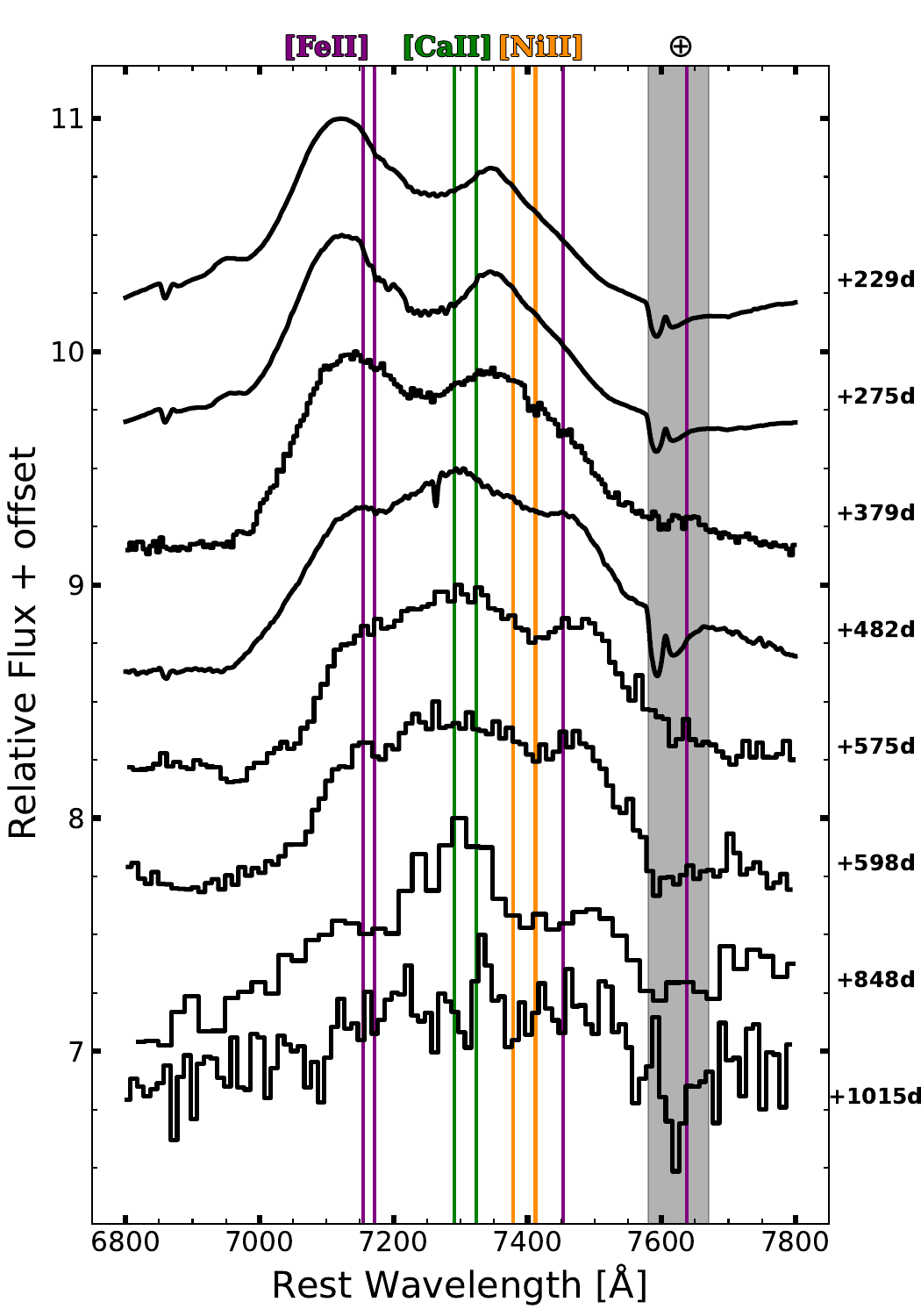}
    \caption{Temporal evolution of the $7300$~\AAA feature, transitioning from \fFe{2}+\fNi{2} to \CaII around $\sim 480$~days after \tmax (see \S\ref{subsec:results.caii}).}
    \label{fig:CaIIannotate}
\end{figure}

\begin{figure}
    \centering
    \includegraphics[width=\linewidth]{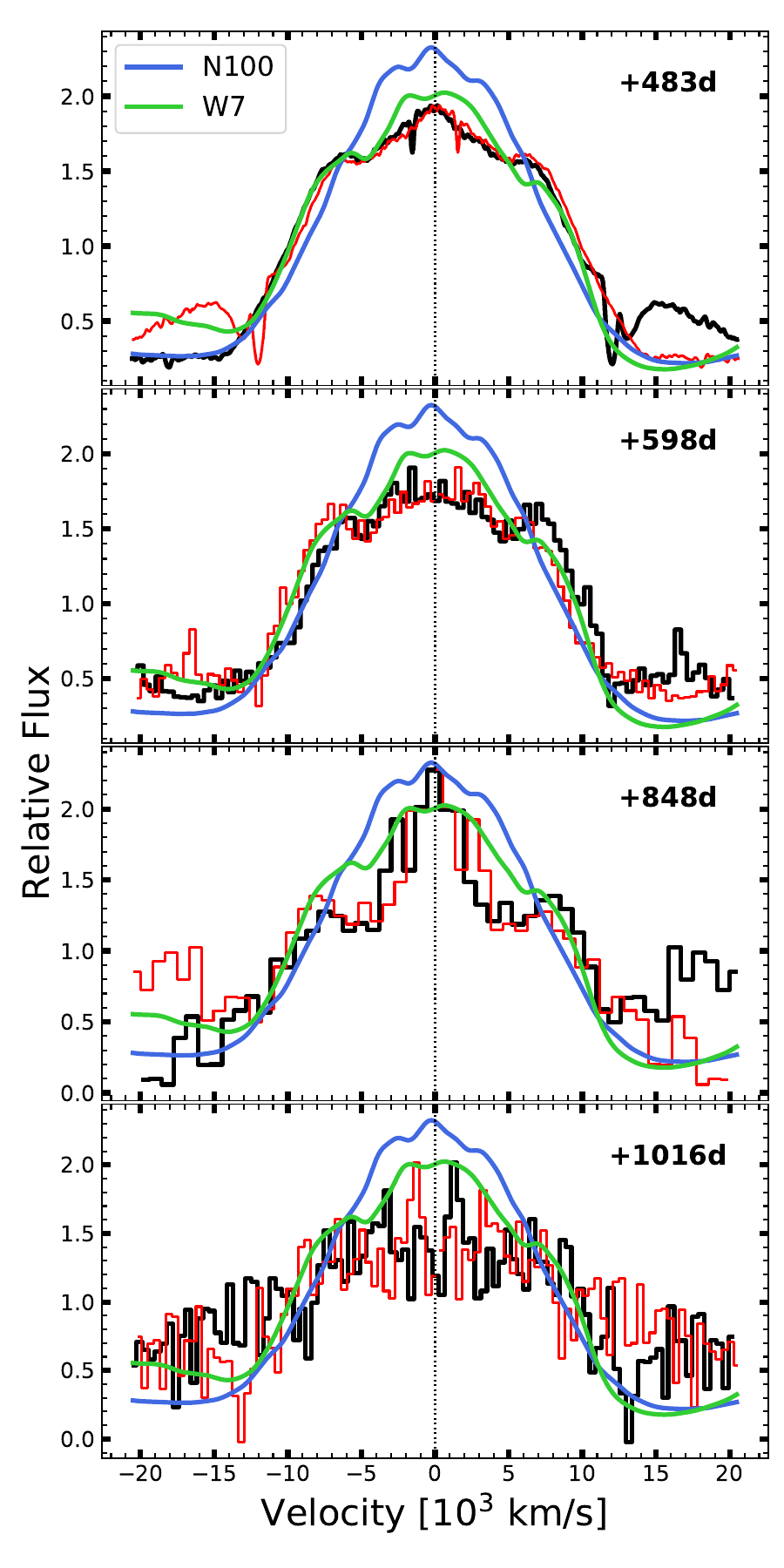}
    \caption{Velocity evolution of the continuum-subtracted \CaII profile at $>450$~days (black) compared to the theoretical 1000-day nebular spectra computed by \citet{fransson2015} using density profiles from the W7 deflagration model \citep[green, ][]{iwamoto1999} and the N100 delayed-detonation model \citep[blue, ][]{seitenzahl2013b}. The thin red lines are the line profile reflected across the line center to highlight the profile symmetry. }
    \label{fig:CaIIfold}
\end{figure}

\begin{figure}
    \centering
    \includegraphics[width=\linewidth]{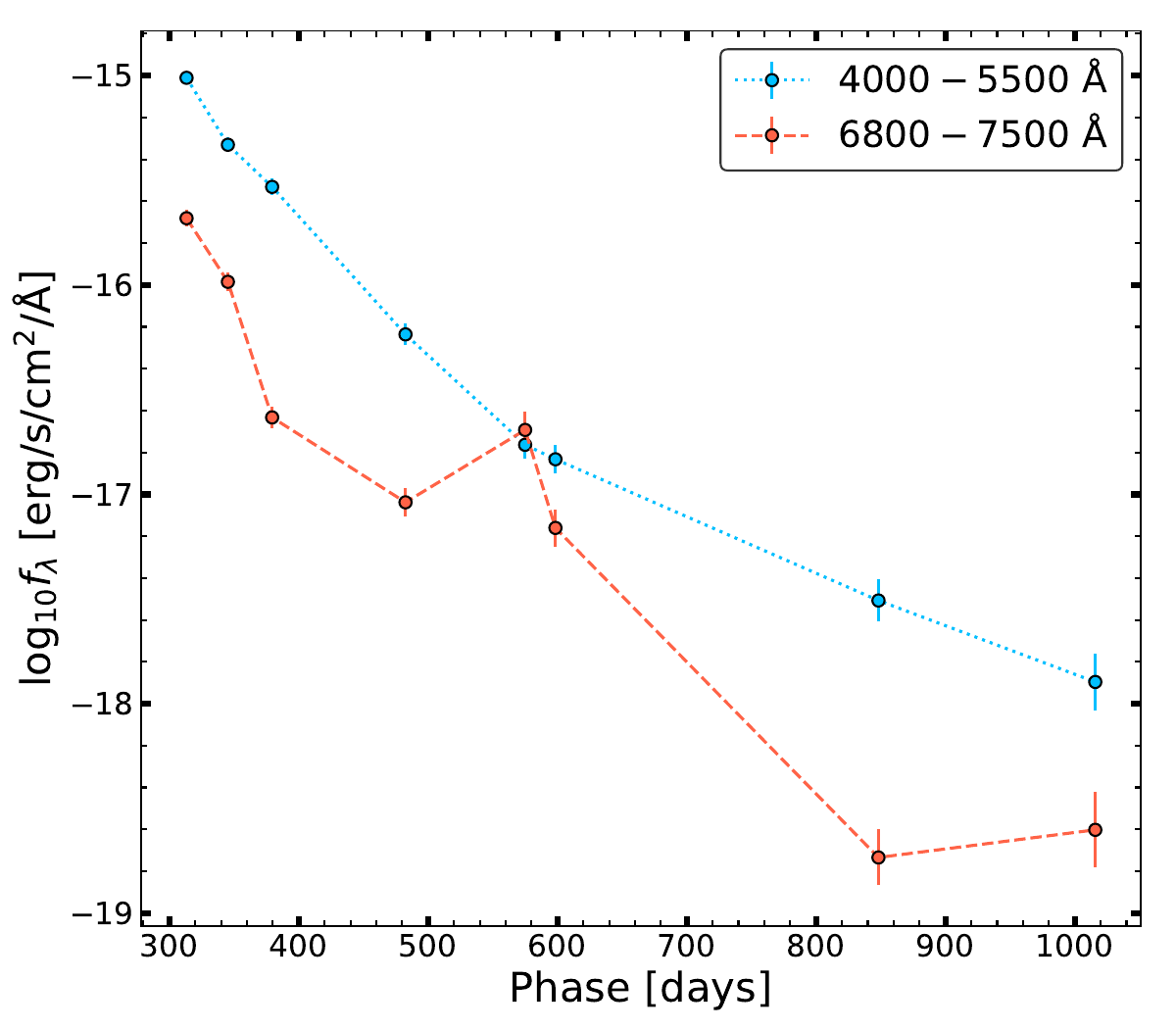}
    \caption{Evolution of the average flux for the Fe-dominated $4000-5500$~\AAA region (dotted blue) compared to the feature at $\approx 7300$~\AAA (dashed red).}
    \label{fig:boxfilter}
\end{figure}

The feature at $\sim 7300$~\AAA is dominated by a blend of \fFe{2} and \fNi{2} in the nebular spectra of normal \sneia \citep[e.g., ][]{mazzali2015, maguire2018, flors2018, wilk2020}. However, Figs. \ref{fig:fluxcal_spectra}, \ref{fig:normspex}, and \ref{fig:CaIIannotate} show a distinct change in this feature at $\approx 480$~days as well, with the emission profile transitioning from double- to triple-peaked. We consider 4 possible interpretations:

\begin{enumerate}
    \item \fFe{2} and \fNi{2} produce the entire profile with no contribution from \CaII;
    \item {}[\ion{Ca}{2}] is narrow ($<4000$~\kms) and only contributes to the central peak, whereas moderate-width ($\approx 8000$~\kms) \fFe{2} and \fNi{2} produce the wings;
    \item Broad \CaII ($\sim 12\,000$~\kms) dominates the feature, with minor contributions from narrow ($\lesssim 2500$~\kms) \fFe{2} and \fNi{2} to the wings; or
    \item The entire profile is produced by \CaII with no significant contribution from \fFe{2}+\fNi{2}.
\end{enumerate}

Case 1, where the central emission peak is attributed to \fFe{2} and/or \fNi{2}, requires a highly-asymmetric ejecta distribution which conflicts with the symmetric emission-line profiles of Co, Ni, and Fe \citep[e.g., ][]{mcclelland2013, graham2015a}. Additionally, the spherically-symmetric, one-dimensional models of \citet{fransson2015} do not require any ejecta asymmetry to reproduce the observations. Thus, we conclude the central peak is rest-velocity \CaII.

Case 2, where \fFe{2} and \fNi{2} dominate the profile with only a minor contribution from \CaII, is plausible but we find it unlikely for several reasons. First, it requires \CaII to be present in the ejecta but only at low velocities ($\lesssim 4000$~\kms) that are typically dominated by iron-group elements \citep[IGEs; e.g., ][]{ruizlapuente1992, liu1997}. Second, high-density burning can produce $^{48}$Ca \citep{meyer1996, dominguez2000} but it is likely confined to the lowest velocities \cite[$\lesssim 1000$~\kms; e.g, ][]{galbany2019}, which disagrees with the observed $\approx 4\,000$~\kms profile. Finally, $^{40}$Ca is readily produced by incomplete Si burning \citep[e.g., ][]{thielemann1986} and Ca is observed in spectra of \name near maximum light at $(10-25)\times 10^3$~\kms \citep[e.g., ][]{parrent2012, pereira2013, zhang2016}. Thus, attributing only the central component to \CaII would require two distinct, non-overlapping zones within the ejecta: an outer, high-velocity zone ($v > 10\,000~\rm{km}~\rm s^{-1}$) responsible for the Ca near maximum light and an inner, low-velocity zone ($v < 5,000~\rm{km}~\rm s ^{-1}$) responsible for the nebular-phase Ca emission. This scenario is unlikely as it disagrees with the chemical stratification resulting from nuclear burning \citep[e.g., ][]{nomoto1984, thielemann1986}. 

Case 3, where the profile is dominated by high-velocity \CaII with minor contributions from \fFe{2} and \fNi{2}, is more likely considering our knowledge of chemical distribution and stratification in the ejecta \citep[e.g., ][]{wilk2020}. This interpretation satisfies the requirement for high-velocity Ca without requiring the \fFe{2} and \fNi{2} contributions to disappear completely. However, attempting to model this feature with a high-velocity Ca component and Fe+Ni contributing to the wings does not provide a satisfactory fit. The residual peaks after removing a broad ($\approx 12\,000$~\kms) \CaII component do not have a self-consistent velocity shift for the \fFe{2} and \fNi{2} line profiles. The blue-shifted peak can be attributed to slightly blue-shifted ($\approx -500~\rm{km}~\rm s^{-1}$) \fFe{2}, similar to $<400$~day spectra \citep[e.g., ][]{mcclelland2013}, but it then requires red-shifted \fFe{2} ($\approx +1500~\rm{km}~\rm s^{-1}$) or \fNi{2} ($\approx + 3500~\rm{km}~\rm{s}^{-1}$) to explain the peak at $\approx 7500$~\AAA. These mismatching velocity shifts suggest this is also an unsatisfactory explanation for the $7300$~\AAA feature. 

This leaves Case 4, where both the central peak and wing components are all created by \CaII. To test this idea, we show the line profile reflected about the central wavelength in Fig. \ref{fig:CaIIfold}. Allowing for some ambiguities in the central wavelength because it is a blended doublet, the profiles are remarkably symmetric. More evidence for the \CaII interpretation comes from the lack of significant evolution between the $\approx 480$-day spectrum and the $\approx 1000$-day spectrum. The $4000-6000$~\AAA region shows noticeable evolution in both line profiles and line strengths over these epochs, likely due to the dominant line emission shifting from \pFe{3} to \pFe{2}+\pFe{1} (\S\ref{subsec:results.feblend}). We would expect some evolution at $(t-t_{\rm{max}})\gtrsim 500$~days in the $7300$~\AAA profile if \fFe{2} and \fNi{2} contributed significantly to the emission profile. 

Thus, we attribute the feature at $\sim 7300$~\AAA to \CaII at $\gtrsim 450$~days after \tmax. However, this requires a rapid ($\lesssim 100$~days) shift in the ionization state of the ejecta from $\sim 450-550$~days after explosion and has interesting implications for the temperature and density evolution of the ejecta, as \fFe{3} disappears at roughly the same epoch that 
\CaII appears (\S\ref{subsec:results.feblend}). 

This spectral evolution cannot be attributed to weak features becoming visible as other features fade. Fig. \ref{fig:boxfilter} compares the average flux of the Fe-dominated $4000-6000$~\AAA region and the $\sim 7300$~\AAA feature,\footnote{\input{footnotes/boxfilter}} showing that the Fe blends evolve smoothly with time albeit with a distinct change in the decay rate at $\approx 550$~days after \tmax. The \CaII feature generally exhibits a similar decline in flux but shows strong variations, especially at $\approx 500-600$~days after \tmax when the Fe-region changes its decline rate. This general time period agrees with the end of the nebular-phase NIR plateau discovered by \citet{graur2020} and coincides with a change in the optical-to-NIR flux ratio \citep{maguire2016, dimitriadis2017, graur2020}, all of which suggest a distinct shift in emission properties.

Interestingly, the $1000$-day theoretical nebular spectra computed by \citet{fransson2015} using delayed-detonation \citep[N100; ][]{seitenzahl2013} and pure deflagration \citep[W7; ][]{iwamoto1999} explosion models both predict symmetric, triple-peaked \CaII profiles which qualitatively match those shown in Fig. \ref{fig:CaIIfold}. However, the 1000-day model spectra are most similar to the $\leq 600$-day observed spectra, highlighting the need for additional modeling at these epochs.

\section{Progenitor System and Explosion Mechanism Constraints}\label{sec:progenitor}

\begin{figure}
    \centering
    \includegraphics[width=\linewidth]{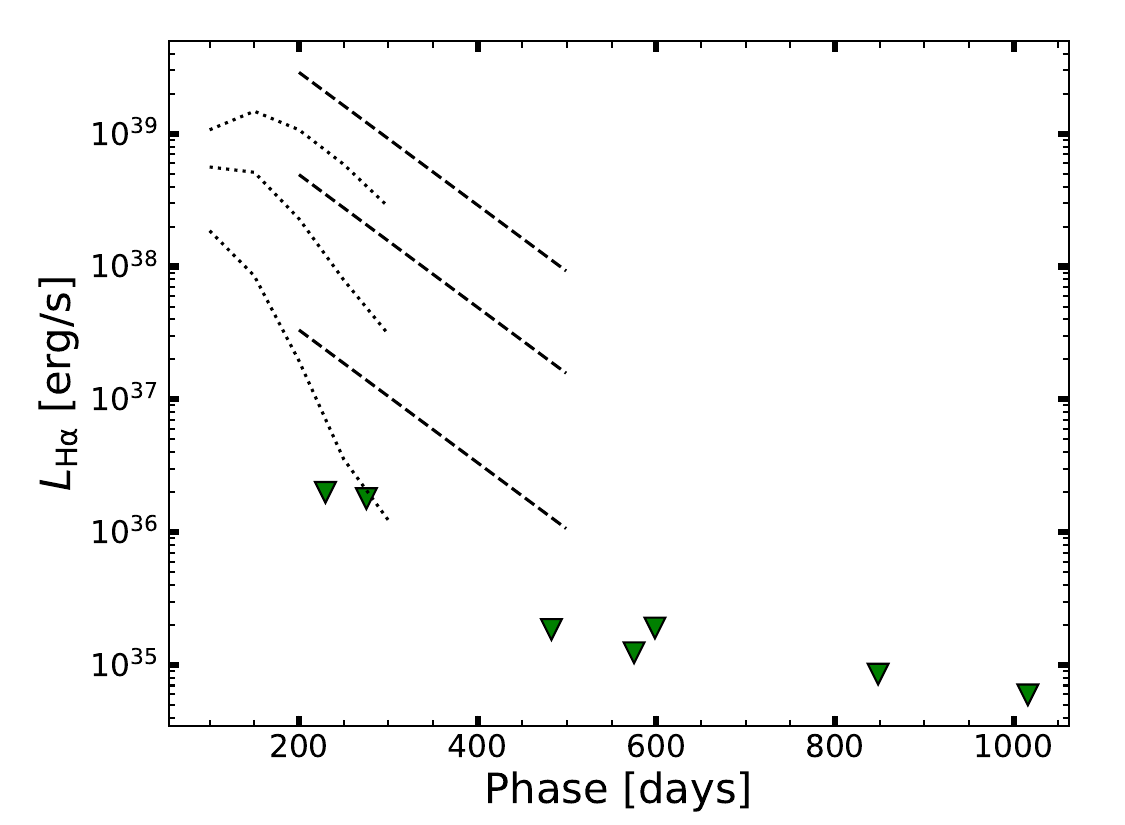}
    \caption{Non-detections of \Ha (inverted triangles) compared to the models of \citet[][dotted lines, $100\, \rm d \lesssim t_{\rm{max}} \lesssim 300\,\rm d$]{dessart2020} and \citet[][dashed lines, $200\,\rm d \lesssim t_{\rm{max}} \lesssim 500\,\rm d$]{botyanszki2018} for ablated masses of $(10^{-1}, 10^{-2}, 10^{-3})~\rm M_\odot$, from top to bottom.} Flux limits for \Ha, \ion{He}{1}$\lambda5876$, and [\ion{O}{1}]$\lambda6300$ are included as supplementary material.
    \label{fig:Hlims}
\end{figure}

The nebular spectra also provide constraints on the progenitor system of \name. We discuss scenarios involving mass-transfer from non-degenerate companions (i.e., the ``single-degenerate'' scenario) in \S\ref{subsec:progenitor.single} and discuss double WD systems (i.e., the ``double-degenerate'' scenario) in \ref{subsec:progenitor.double}.

\subsection{Single-Degenerate Scenarios}\label{subsec:progenitor.single}

If a non-degenerate star was undergoing Roche Lobe overflow (RLOF) and depositing mass onto the WD at time of explosion, numerical simulations predict ${0.1-0.5~ \rm M_\odot}$ of mass will be unbound from the stellar envelope \citep[e.g., ][]{marietta2000, pan2012a, boehner2017}, depending on the type of donor star (i.e, main-sequence versus red-giant stars). Additionally, the surrounding environment may contain H-rich material from the companion star wind or from nova-like eruptions on the WD surface during the accretion phase \citep[e.g, ][]{hamuy2003, walder2008, moore2012}. Signatures for such material have been searched for previously in \name \citep{shappee2013, lundqvist2015, graham2015b}, without success. 

Neither H nor He are observed at any point in the spectra of \name and we place $10\sigma$ non-detection limits on emission lines from \Ha and \ion{He}{1}$\lambda5876$ in the flux-calibrated spectra using Eq. 3 from \citet{tucker2020}. We assume a line width of $1000~\rm{km}~\rm s^{-1}$ which is predicted for stripped companion material \citep[e.g., ][]{boehner2017} and in rough agreement with emission-line widths seen in SNe Ia interacting with nearby dense CSM ($v = 500-2000~\rm{km}~\rm s^{-1}$; \citealp{silverman2013, silverman2013b, graham2019}). The flux non-detections scale with the assumed velocity as $\propto \sqrt{ v / 1000 ~ \rm{km}~\rm{s}^{-1} }$. Our adopted 10$\sigma$ flux limit is conservative but corresponds to a line profile that would be visibly obvious.

Fig. \ref{fig:Hlims} compares the \Ha non-detections, including the flux calibration uncertainties, to the most recent models \citep{botyanszki2018, dessart2020}. \citet{botyanszki2018} do not provide time-dependent line luminosities so we adopt the time-dependence derived by \citet{tucker2020}. The limit on any potential stripped/ablated companion material is $\lesssim 10^{-3}~M_\odot$ for both models, essentially excluding all non-degenerate H-rich donor stars. The \Ha non-detections also disfavor the presence of an H-rich CSM. Assuming a wind velocity of 10 (100) \kms, this precludes any significant CSM production by the progenitor system in the past 7000 (700) years. The lack of CSM is also difficult to reconcile with the core-degenerate scenario \citep{soker2014}, although this interpretation necessarily depends on the time between merger with the stellar core and explosion.

Interpreting the lack of He is more complicated. \citet{botyanszki2018} provide a simplified He-star model which replaces the H-rich material unbound from a main-sequence (MS) star with He-rich material. This simple model provides a direct estimate of He line luminosities but fails to capture the physical conditions of a He-star envelope. The differing density profiles and surface gravities between MS and He-star envelopes will affect both the amount of unbound mass and its velocity. Additionally, the models of \citet{dessart2020} include time-dependent opacities and predict that only the NIR \ion{He}{1}~$1.083~\mu\rm m$ is produced at detectable levels. Thus, while we can confidently exclude any He emission at similar flux limits to those on \Ha shown in Fig. \ref{fig:Hlims}, it is not clear how to interpret this as a limit on ablated mass from a companion.

\subsection{Double-Degenerate Scenarios} \label{subsec:progenitor.double}

Observational signatures of double-degenerate progenitor systems are subtle due the compact nature of both stars, but some constraints can be obtained from the nebular-phase spectra. First, we checked for [\ion{O}{1}]$\lambda\lambda6300,6364$ emission lines as this may indicate the violent merger of two WDs \citep{kromer2013} as seen in the subluminous SN Ia 2010lp \citep{taubenberger2013}. However, both the archival and new spectra have no evidence for any O emission. This argues against a violent merger producing \name, but the nebular-phase model parameter space for violent mergers is also largely unexplored. 

Another potential double-degenerate scenario is a direct (head-on) collision, usually induced by orbital perturbations from external bodies \citep[e.g., ][]{thompson2011, antognini2014}, which then produces highly-asymmetric ejecta \citep[e.g., ][]{rosswog2009, vanrossum2016}. Observationally, bimodal and asymmetric emission-line profiles of iron-group elements (IGEs) have been seen in nebular-phase \sneia spectra and used to infer explosion asymmetry \citep[e.g., ][]{dong2015, mazzali2018, vallely2020, hoeflich2021}. However, the Co, Fe, and \CaII line profiles are symmetric (see \S\ref{sec:results}) and the one-dimensional models of \citet{fransson2015} do not require any asymmetry to reproduce the observations (Fig. \ref{fig:CaIIfold}). Thus, we find no evidence for explosion asymmetry which is tentative evidence against the direct-collision interpretation, in agreement with the inferred nucleosynthetic yield \citep{tucker2021c}.

\section{Discussion}\label{sec:discuss}

Our new spectra reveal a hitherto unseen transition in the ejecta $\sim 500$~days after explosion. This epoch also corresponds to the decades-old prediction of \citet{axelrod1980} who proposed that an infrared catastrophe occurs when the ejecta temperature and density are too low to populate the $\sim 3~\rm{eV}$ \fFe{3} transitions. However, \citet{axelrod1980} predicted that \textit{all} emission will be shifted into the mid-IR (MIR) regime as the only remaining cooling mechanism would be by exciting fine-structure transitions. 

High-energy ($\sim 1~\rm{MeV}$) \iCo{56} positrons dominate the energy input at $\lesssim 1200$~days explosion \citep[e.g., ][]{tucker2021c}. Some of these positrons heat the ejecta directly but the majority of the positrons produce non-thermal excitations and ionizations (e.g., \citealp{kozma1992, jerkstrand2011, li2012, shingles2020}), either directly or via downgraded electrons. UV photons are produced upon recombination which are transferred to optical and NIR wavelengths though multiple scatterings and fluorescence due to the optically-thick forest of Fe transitions in the UV \citep[e.g., ][]{pinto2000, jerkstrand2011, fransson2015}. This scenario is consistent with the failure to detect \name at $< 4000$~\AAA during the nebular phase \citep{kerzendorf2017}. 

Non-local radiative transfer effects explain the continued presence of optical emission at these epochs \citep{fransson2015}, but it does not account for the observed change in emission properties at $\sim 500$~days after explosion. The shift from \ion{Fe}{2}+\ion{Fe}{3} to \ion{Fe}{1}+\ion{Fe}{2} suggests a ``recombination wave'' propagating through the \iNi{56} region \citep{graur2020}. However, we observe the transition at $\sim 500$~days after explosion rather than $\sim 600$~days predicted by the models of \citet[][see their Fig. 1]{fransson2015}. This discrepancy may seem minor but for spherically-symmetric expansion, this changes the density at the transition by a factor of $\approx 1.5-2$ which has important ramifications for energy deposition \citep[c.f., ][]{axelrod1980}. 

We propose that clumping may explain this discrepancy. Clumping in the ejecta of \sneia has been suggested by both observational \citep[e.g., ][]{black2016, mazzali2020} and theoretical \citep[e.g., ][]{wilk2020} studies, but confirmation is difficult. Clumping determines where energy deposition occurs as clumps will retain energy input from radioactive decays more efficiently than the lower-density regions between them. However, the higher electron density in the clumps also increases the recombination rate and lowers the average ionization state \citep[e.g., ][]{mazzali2020}. \citet{wilk2020} show that varying the level of clumping, parameterized as a ``filling factor'', has a profound impact on the observed spectra. Increased clumping diminishes the strong \fFe{3} blend at $\sim 4600$~\AAA while increasing the strength of the \CaII lines. Both signatures qualitatively match the spectral transition we observe $\sim 500$~days after explosion. 

However, it is unclear if these simple comparisons match the true physical evolution of the ejecta, as the interpretation of \CaII is also dependent on the adopted explosion model. The \Mch and sub-\Mch explosions models from \citet{wilk2020} differ in Ca production by a factor of $\sim 2$. The sub-\Mch double-detonation explosion models of \citet{polin2021} also predict nebular-phase \CaII. The off-center \Mch delayed-detonation models of \citet{hoeflich2021}, designed to replicate a low-luminosity \snia, predict strong and asymmetric nebular-phase \CaII emission. Thus, it remains unclear which model(s) accurately predict \CaII emission after the nebular-phase ionization change without introducing new discrepancies with other observations of \name.  

Observationally, \CaII is absent in spectra of \sneia obtained $\lesssim 500$~days after explosion \citep[e.g., ][]{graham2017, maguire2018, flors2018, flors2020}. However, \CaII is observed in the underluminous 91bg-like (e.g., SNe 1991bg, \citealp{turatto1996}; 1999by, \citealp{blondin2018}; 2006mr, \citealp{stritzinger2010}) and 02es-like (e.g., SNe 2010lp, \citealp{taubenberger2013}; 2019yvq, \citealp{siebert2020,tucker2021a,burke2021}) subclasses of \sneia at $\lesssim 200$~days after explosion. Considering that \CaII is an effective coolant and a resonant transition, it is perhaps unsurprising that the time-dependence of \CaII emission is related to the temperature and density in the ejecta. However, the \CaII lines in the underluminous \sneia are also flat-topped and symmetric, similar to \name (Fig. \ref{fig:CaIIfold}). If \CaII is indeed a reliable probe of the ejecta geometry, it is interesting that the \CaII line profiles are so similar. However, interpreting symmetric \CaII emission as evidence for symmetric ejecta introduces new tensions with the complex elemental distributions seen in some \snia remnants \citep[e.g., ][]{stone2021}. 

\name presented a rare opportunity to study \sneia physics in unprecedented detail and it remains one of the best-studied astronomical objects to-date \citep[e.g., ][]{tucker2021c}. If the sharp ionization change is related to the infrared catastrophe predicted by \citet{axelrod1980}, this transition should produce a distinct increase in the MIR flux due to fine-structure cooling. Nebular-phase \textit{Spitzer} and \textit{Herschel} observations of \name \citep{johansson2013, johansson2014} did not cover the wavelength range ($20-40~\mu\rm m$; e.g., \citealp{fransson2015}) expected for fine-structure emission lines but the upcoming \textit{James Webb Space Telescope} should provide a direct test of this theory \citep[e.g., ][]{ashall2021}.

\section*{Acknowledgements}

We thank Peter Hoeflich, Zach Claytor, Connor Auge, and Michelle Togami for useful discussions. 

M.A.T. acknowledges support from the DOE CSGF through grant DE-SC0019323. CSK and KZS are supported by NSF grants AST-1814440 and AST-1907570.

The LBT is an international collaboration among institutions in the
United States, Italy and Germany. LBT Corporation partners are: The
University of Arizona on behalf of the Arizona Board of Regents;
Istituto Nazionale di Astrofisica, Italy; LBT Beteiligungsgesellschaft,
Germany, representing the Max-Planck Society, The Leibniz Institute for
Astrophysics Potsdam, and Heidelberg University; The Ohio State
University, representing OSU, University of Notre Dame, University of
Minnesota and University of Virginia.

\bibliography{ref}{}
\bibliographystyle{aasjournal}



\end{document}

%% file: def.tex
\usepackage{xspace, comment, amsmath}

\newcommand{\name}{SN~2011fe\xspace}
\newcommand{\tmax}{$t_{\rm{max}}$\xspace}

\newcommand{\Mch}{$\rm M_{\rm{Ch}}$\xspace}

\newcommand{\Ha}{H$\alpha$\xspace}

\newcommand{\CaII}{[\ion{Ca}{2}]\xspace}

\newcommand{\fFe}[1]{[\ion{Fe}{#1}]\xspace}
\newcommand{\pFe}[1]{\ion{Fe}{#1}\xspace}
\newcommand{\fNi}[1]{[\ion{Ni}{#1}]\xspace}
\newcommand{\fCo}[1]{[\ion{Co}{#1}]\xspace}

\newcommand{\iNi}[1]{$^{#1}$Ni\xspace}
\newcommand{\iCo}[1]{$^{#1}$Co\xspace}

\newcommand{\M}[1]{$M_{#1}$\xspace}

\newcommand{\snia}{SN Ia\xspace}
\newcommand{\sneia}{SNe Ia\xspace}

\newcommand{\kms}{km~s$^{-1}$\xspace}

\newcommand{\AAA}{\AA\xspace}


%% file: lbtinfo.tex
\begin{deluxetable}{lccr}
\tablecaption{New LBT MODS Spectroscopy. \label{tab:lbtinfo}}
\tablehead{
    \colhead{UT Date} & \colhead{Phase\tablenotemark{a}} & \colhead{Exp. Time} \\ 
     & [days] & [s]
}
\startdata
2013-01-05 & 482.6 & 14650 \\
2013-05-01 & 598.4 & 10800 \\
2014-01-06 & 848.2 & 10800 \\
\enddata
\tablenotetext{a}{Relative to $t_{\rm{max}}$.}
\end{deluxetable}

%% file: footnotes/boxfilter.tex
The average flux for the $\sim 7300$~\AAA region is cut off at 7500~\AAA to prevent the O$_2$ telluric band from affecting the results (e.g., Fig. \ref{fig:CaIIannotate}).